\documentclass{aa}
\usepackage{txfonts}
\usepackage{graphicx}
\usepackage{natbib}

\bibpunct{(}{)}{;}{a}{}{,}

\newcommand{\oam}{\mbox{\object{AM\,Her}}}
\newcommand{\amher}{AM\,Her}
\newcommand{\amherc}{AM\,Herculis}
\newcommand{\ten}[2]{#1\times 10^{#2}}

\newcommand{\msunyr}{M$_\odot\,\mathrm{yr}^{-1}$}
\newcommand{\msun}{M$_\odot$}

\newcommand{\teff}{$T_\mathrm{eff}$}

\newcommand{\nh}{$N_\mathrm{HI}$}

\newcommand{\atoms}{H-atoms\,cm$^{-2}$}

\newcommand{\erg}{erg\,s$^{-1}$}
\newcommand{\ergs}{erg\,cm$^{-2}$s$^{-1}$}
\newcommand{\ergsa}{erg\,cm$^{-2}$s$^{-1}$\AA$^{-1}$}

\newcommand{\gs}{g\,s$^{-1}$}
\newcommand{\lya}{Ly$\alpha$}

\newcommand{\tbb}{$T_{\rm bb}$}
\newcommand{\ktbb}{k$T_{\rm bb}$}

\newcommand{\chisq}{$\chi^2$}
\newcommand{\chisqr}{$\chi^2_\nu$}

\newcommand{\rosat}{\textit{ROSAT}}
\newcommand{\chandra}{\textit{Chandra}}
\newcommand{\euve}{\textit{EUVE}}

\begin{document}

\title{Soft X-ray spectral variability of \amherc\thanks{Based on
observations made with the \rosat\ satellite.}}

\author {K.\ Beuermann \inst{1}
\and     E. El Kholy{\inst{2,1}}     
\and     K. Reinsch \inst{1}}


\institute{Institut f\"ur Astrophysik, Friedrich-Hund-Platz 1,
D-37077 G\"ottingen, Germany, beuermann or reinsch@astro.physik.uni-goettingen.de,
\and
National Research Institute of Astronomy and Geophysics, Helwan,
Cairo, Egypt, essam@nriag.sci.eg}

\date{Received December 28, 2007 /  February 12, 2008}
 
\authorrunning{K. Beuermann et al.}  \titlerunning{Soft X-ray spectral
variability of AM Herculis}

\abstract { 
Polars (\amherc\ binaries) are a prominent class of bright soft X-ray
sources, many of which were discovered with \rosat.}
{ 
We present a homogenous analysis of all the pointed \rosat\ PSPC
observations of polars subdivided into two papers that discuss
the prototype polar \amher\ in detail and summarize the class
properties of all other polars.}
{ 
We derive the high-state soft X-ray flux and short-term spectral
variability of \amher\ using a new detector response matrix and a
confirmed flux calibration of the \rosat\ PSPC below 0.28\,keV. }
{ 
The best-fit mean single-blackbody temperature and integrated
bright-phase energy flux of \amher\ in its April 1991 high state are
$27.2\pm1.0$\,eV and $\ten{(2.6\pm0.6)}{-9}$\,\ergs, respectively. The
total blackbody flux of a multi-temperature model that fits both the soft
X-ray \emph{and} the fluctuating far-ultraviolet components is
$F_\mathrm{bb}=\ten{(4.5\pm1.5)}{-9}$\,\ergs.  The total accretion
luminosity at a distance of 80\,pc,
$L_\mathrm{bb}=\ten{(2.1\pm0.7)}{33}$\,\erg, implies an accretion rate of
$\dot M=\ten{(2.4\pm0.8)}{-10}$\,\msunyr\ for an 0.78\,\msun\ white
dwarf. The soft X-ray flux displays significant variability on time
scales down to 200\,ms. Correlated spectral and count-rate variations
are seen in flares on time scales down to 1\,s, demonstrating the
heating and cooling associated with individual accretion events. }
{ 
Our spectral and temporal analysis provides direct evidence for the
blobby accretion model and suggests a connection between the soft
X-ray and the fluctuating far-ultraviolet components.}

\maketitle

\section{Introduction}

Research on polars, the synchronized magnetic variety of cataclysmic
variables (CVs), has greatly benefited from the discovery of numerous
new members of the class during the \rosat\ All Sky Survey and the
subsequent pointed observations with the position sensitive
proportional counter (PSPC) \citep{beuermannthomas93}. As a
prominent example, we discuss the statistically highly significant
PSPC spectrum of the prototype polar \oam\ and its short-term spectral
variability.  The second paper of this short series of two summarizes
the spectral analysis of all other polars observed with the \rosat\
PSPC.  Before embarking on this project, we have investigated the
putative absolute miscalibration of the PSPC by up to a factor of two
at photon energies below the interstellar carbon edge at
0.28\,keV
\citep{napiwotzkietal93,jordanetal94,wolffetal96,wolffetal99}, where
most of the energy flux of polars resides. We use the new
in-flight calibration of the PSPC for such low-energy photons
\citep{beuermannetal06,beuermann08} and a corrected detector response
matrix of the PSPC that removes defects in the spectra of very soft
sources and leads to improved spectral-fit parameters \citep{beuermann08}.

\section{\rosat\ observations of \amherc}

\amher\ was observed with the PSPC on 12/13 April 1991 for 11.6\,ks in
its high state and on 15/16 September 1991 for 30.6\,ks in an
intermediate or low state. Other short exposures exist, but are not
considered here.  After the decommissioning of the PSPC, \amher\ was
repeatedly observed with the high-resolution imager (HRI), on 13--17
March 1994 and 8--10 March 1995 during high states and between 26
January and 30 May 1996, when it declined from a high state into a
deep low with zero detectable flux.
In all \rosat\ observations, \amher\ was encountered in its normal
mode of accretion with the pole that points more directly towards the
secondary receiving most or all of the transferred matter. The system
is bright for magnetic phases $\phi\simeq 0.2-1.0$, where $\phi\simeq
0$ refers to the linear polarization pulse that occurs when the
accretion spot crosses the limb of the white dwarf and the line of
sight is perpendicular to the magnetic field direction
\citep{heiseverbunt88}. Table~\ref{tab:log} lists all
observations used in this paper along with the AAVSO visual brightness
estimates.

\begin{table}[t]
\caption{Mean AAVSO magnitudes for the AM~Her observations used in
this paper.  The date is for the start of exposure.  H and L refer to
high and low states, respectively.}
\label{tab:log}

\begin{tabular}{l@{\hspace{2.0mm}}c@{\hspace{2.0mm}}c@{\hspace{2.0mm}}c@{\hspace{2.0mm}}ll} 
\hline \hline \noalign{\smallskip}
Instrument &  Date  &  Exp (ks) & State & Magnitude$^{\dagger}$ & Ref.\\           
\noalign{\smallskip} \hline
\noalign{\smallskip}
ROSAT PSPC         & 1991-04-12 &\hspace{-2.0mm} 11.6 & H & $13.0\pm0.2$ &(1)\\
\hspace{10.0mm}PSPC& 1991-09-15 &\hspace{-2.0mm} 30.6 & L & $15.0$ &  \\
\hspace{10.0mm}HRI & 1996-01-26 &\hspace{-0.4mm}  2.3 & H & $13.2$ & \\
EUVE               & 1993-09-23 &\hspace{-1.9mm} 74.0 & \hspace{1.6mm}H--& $13.5\pm0.2$ &\\
                   & 1995-03-08 &\hspace{-2.5mm}121.0 & H & $13.1\pm0.2$ &\\
HUT                & 1995-03-09 &\hspace{-0.2mm}  6.7 & H & $13.1\pm0.2$ &(2)\\
Chandra LETG       & 2000-03-09 &\hspace{-1.7mm} 24.0 & H & $13.3\pm0.2$ &(3)\\
\\
\noalign{\smallskip} \hline  \noalign{\smallskip}
\end{tabular}

$^{\dagger}$ The 'magnitude' is the average of the AAVSO readings over
the duration of the observation and the 'error' is their standard
deviation.\\
References: (1) G\"ansicke et al. (1995), (2) Greeley et al. (1999),
(3)~Burwitz~et~al. (2002), Trill (2006). 
\end{table}


\section{Results}
 
The April 1991 high-state PSPC observation of \amher\ was previously
discussed by \citet{ramsayetal96}. We concentrate here on three
aspects not considered so far, the combined discussion of the soft
X-ray and far-ultraviolet spectra, a comparison of the PSPC data with
the higher resolved \chandra\ and \euve\ spectroscopic observations,
and the spectral variability on time scales down to 1\,s in an attempt
to shed light on the heating and cooling processes in the accretion
spot.

\begin{figure}[t]
\includegraphics[width=87mm,angle=0,clip]{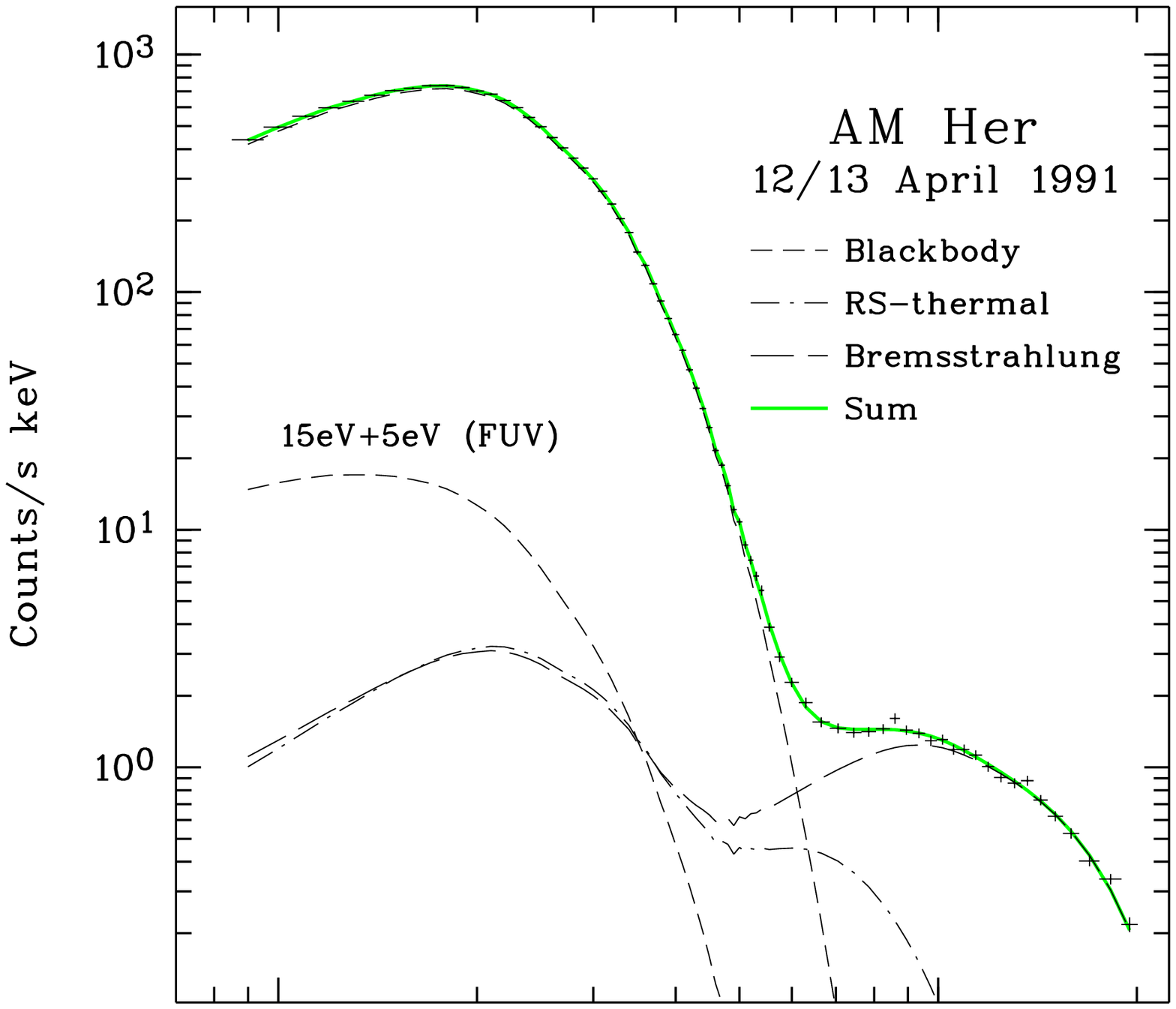}

\vspace*{-0.95mm}
\includegraphics[width=87mm,angle=0,clip]{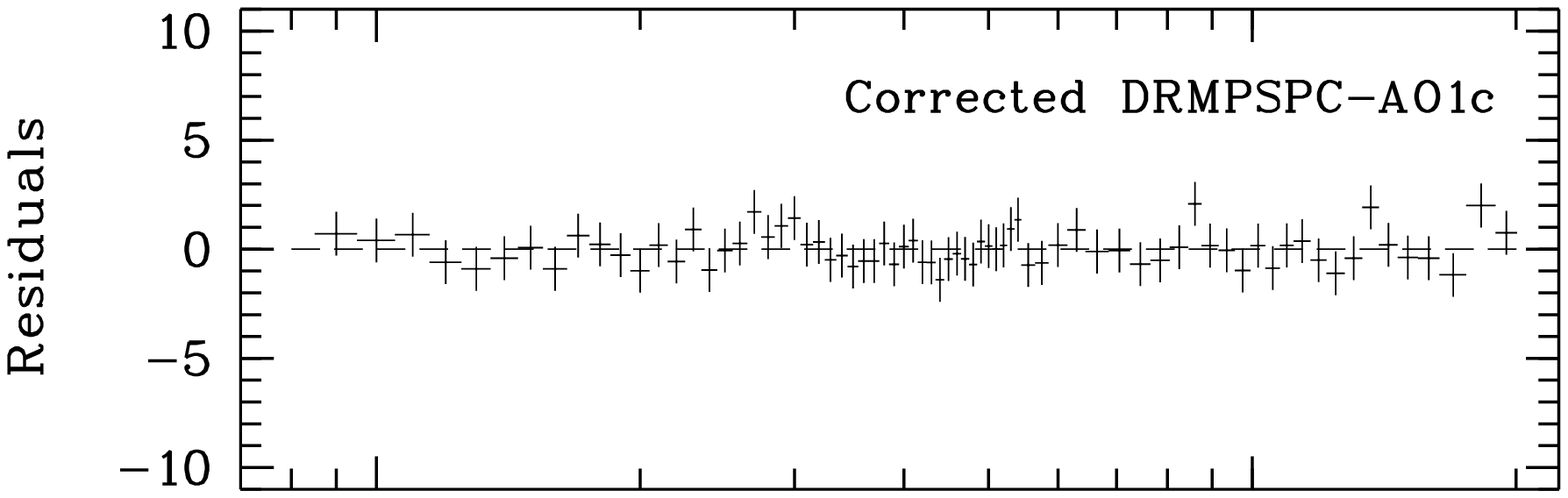}

\vspace*{2mm}
\includegraphics[width=87mm,angle=0,clip]{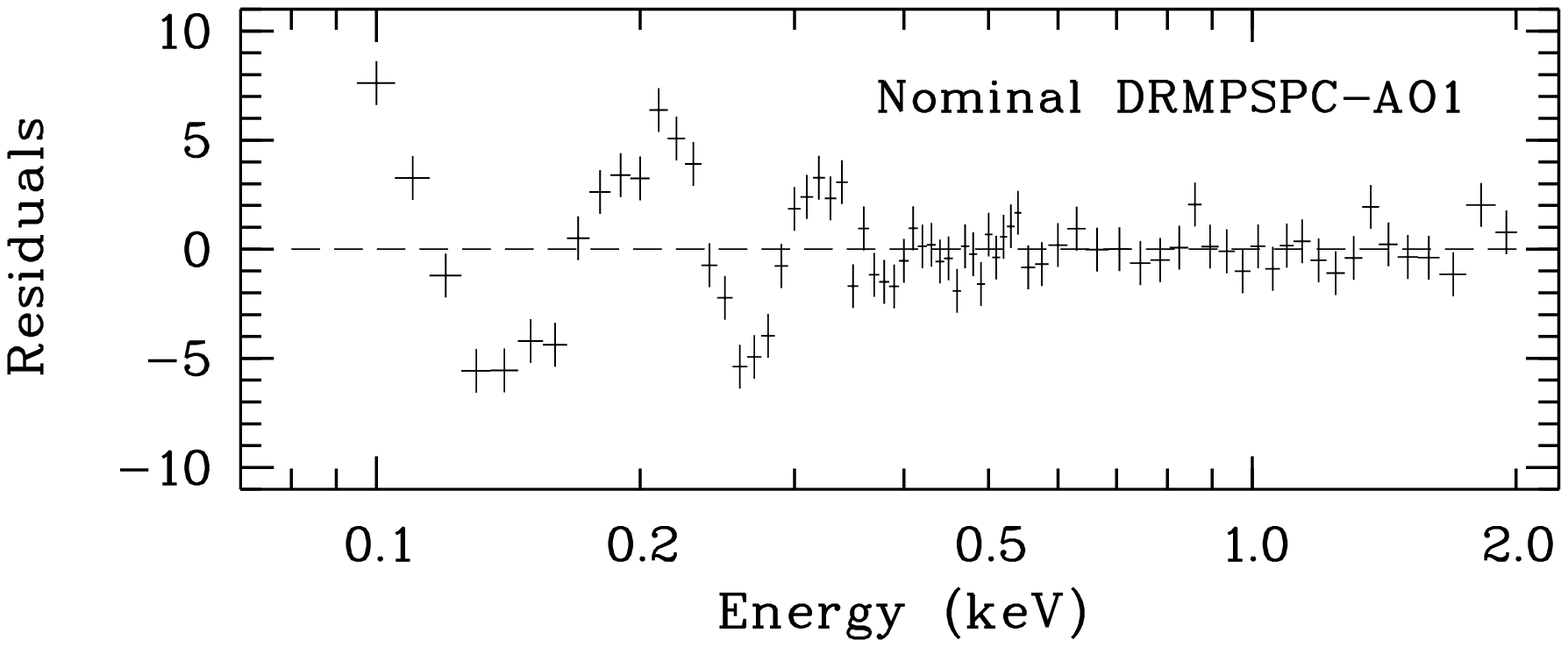}
\caption[chart]{Bright-phase PSPC spectrum of \amher\ in the April
1991 high state. \emph{Top: } Spectral fit of a four-component
spectrum to the data using the corrected detector response matrix
DRMPSPC-AO1c. The dominant blackbody component has
k$T_\mathrm{bb}=27.2$\,eV, the minor components have 15\,eV and 5\,eV.
The bremsstrahlung component has k$T_\mathrm{brems}=20$\,keV and the
low-energy thermal spectrum 0.18\,keV.  \emph{Center and bottom: }
Residuals of the fits obtained with the corrected and the nominal
matrix, respectively. }
\label{fig:meanspec}
\end{figure}

\subsection{Mean bright phase PSPC spectrum of \amher}
\label{sec:brightspec}

We fit the high-state bright-phase spectrum of \amher\ by the sum of
three terms: (i) a thermal bremsstrahlung component with a fixed
temperature of 20\,keV, (ii) a Raymond-Smith thermal spectrum with a
fitted temperature of 0.18\,keV that accounts for the line emission
known to exist at $E<1$\,keV, and (iii) and a quasi-blackbody
component that fits the soft X-ray emission. The actual value of
the bremsstrahlung temperature is not relevant for the fit to the PSPC
spectrum. The integrated bremsstrahlung flux, however, is relevant and the
value quoted below is taken from the fits to the Ginga, ASCA, and RXTE
spectra \citep{beardmoreetal95,ishidaetal97,christian00}.  For
simplicity, we describe interstellar absorption by a column density of
cold gas with solar composition, neglecting both the finite ionization
of the gas and the condensation of metals into dust. In this spirit,
we do not include the ionization edge at 85\AA\ present in the \euve\
and \chandra\ LETG spectra \citep{paerelsetal96,trill06}. Different
assumptions will slightly change the derived blackbody
temperature, but leave the general conclusions unaffected. Our value
of the column density for \amher\ falls within the range of earlier
results \citep{paerelsetal96,ramsayetal96,trill06} and, as those,
exceeds the interstellar \lya-derived atomic hydrogen column density
$N_\mathrm{Ly\alpha}=\ten{(3.0\pm1.5)}{19}$\,\atoms\
\citep{gaensickeetal95}, suggesting additional intra-binary
absorption.

A fit to the mean bright-phase ($\phi=0.22-0.86$) \rosat\ PSPC
spectrum using the above model and the nominal detector response
matrix DRMPSPC-AO1 yields excessive residuals in channels 10 to 40
($E=0.10-0.40$\,keV) with peak and rms values of $7.6\,\sigma$ and
$3.6\,\sigma$, respectively (Fig.~\ref{fig:meanspec}, bottom
panel). These residuals largely disappear with the revised matrix
DRMPSPC-AO1c derived by \citet{beuermann08} and adopted here
(Fig.~\ref{fig:meanspec}, center panel) and the reduced \chisq\ drops
from an unacceptable \chisqr=6.57 to 0.63 (channels 9 to 40),
substantially increasing the confidence in the fit. The best-fit
single-blackbody temperature, cold-matter column density, and total
blackbody flux are k$T_\mathrm{bb,1}=27.3\pm 1.0$\,eV,
$N_\mathrm{H}=\ten{(6.3\pm0.8)}{19}$\,\atoms, and
$F_\mathrm{bb,1}=\ten{(2.6\pm0.2)}{-9}$\,\ergs, respectively, where
the errors refer to the 99\% confidence limit. Since individual
sections of the soft X-ray emitting spot heat and cool independently,
the one-temperature fit must be considered a first approximation and
the total flux may differ if a range of temperatures is assumed. We
consider adding components with a higher and lower temperature in
turn. The addition of a blackbody fixed at k$T_\mathrm{bb,2}=35$\,eV
(see Sect.~\ref{sec:chandra}) lowers k$T_\mathrm{bb,1}$ and raises
$F_\mathrm{bb,1}$, but raises $\chi^2$, too. At the 95\% confidence
level, k$T_\mathrm{bb,1}>22$\,eV and
$F_\mathrm{bb,1}<\ten{5}{-9}$\,\ergs. Adding components with lower
temperatures requires a discussion of the limit that can be placed on
the soft X-ray contribution to the observed ultraviolet flux.

\subsection{The ultraviolet limit}

\begin{figure}[t]
\includegraphics[width=87mm,angle=0]{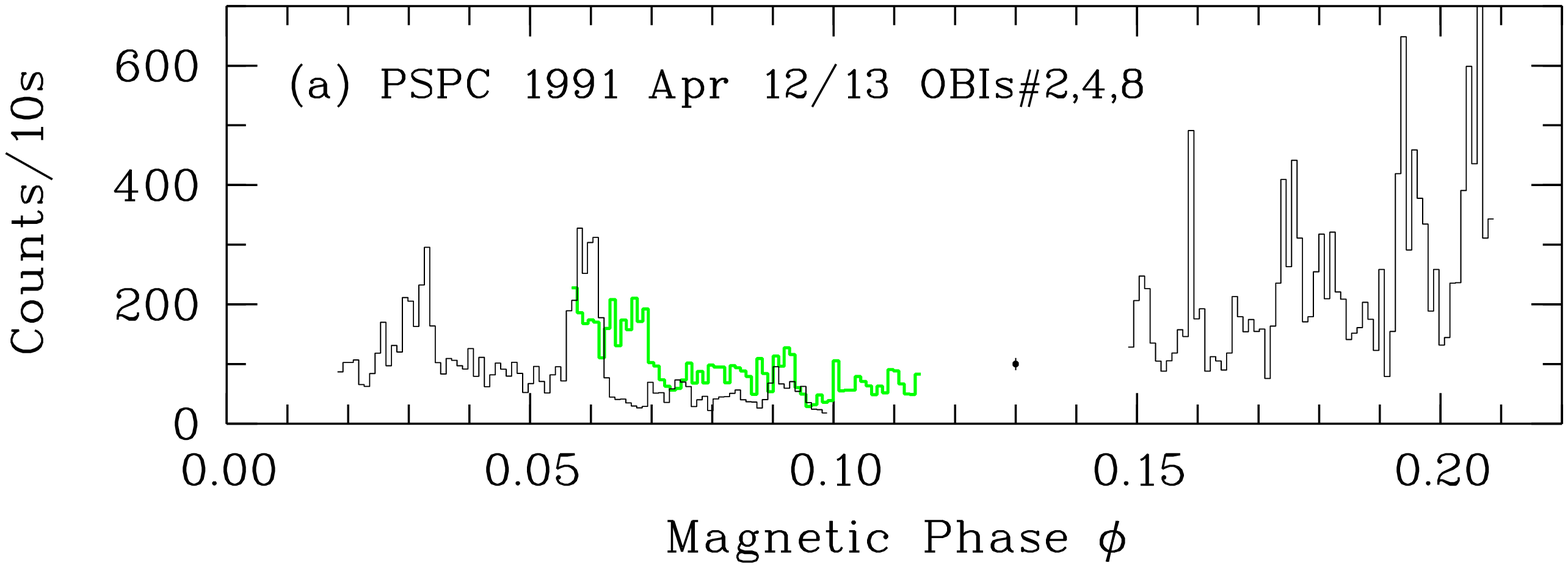}

\vspace*{2mm}\includegraphics[width=87mm,angle=0]{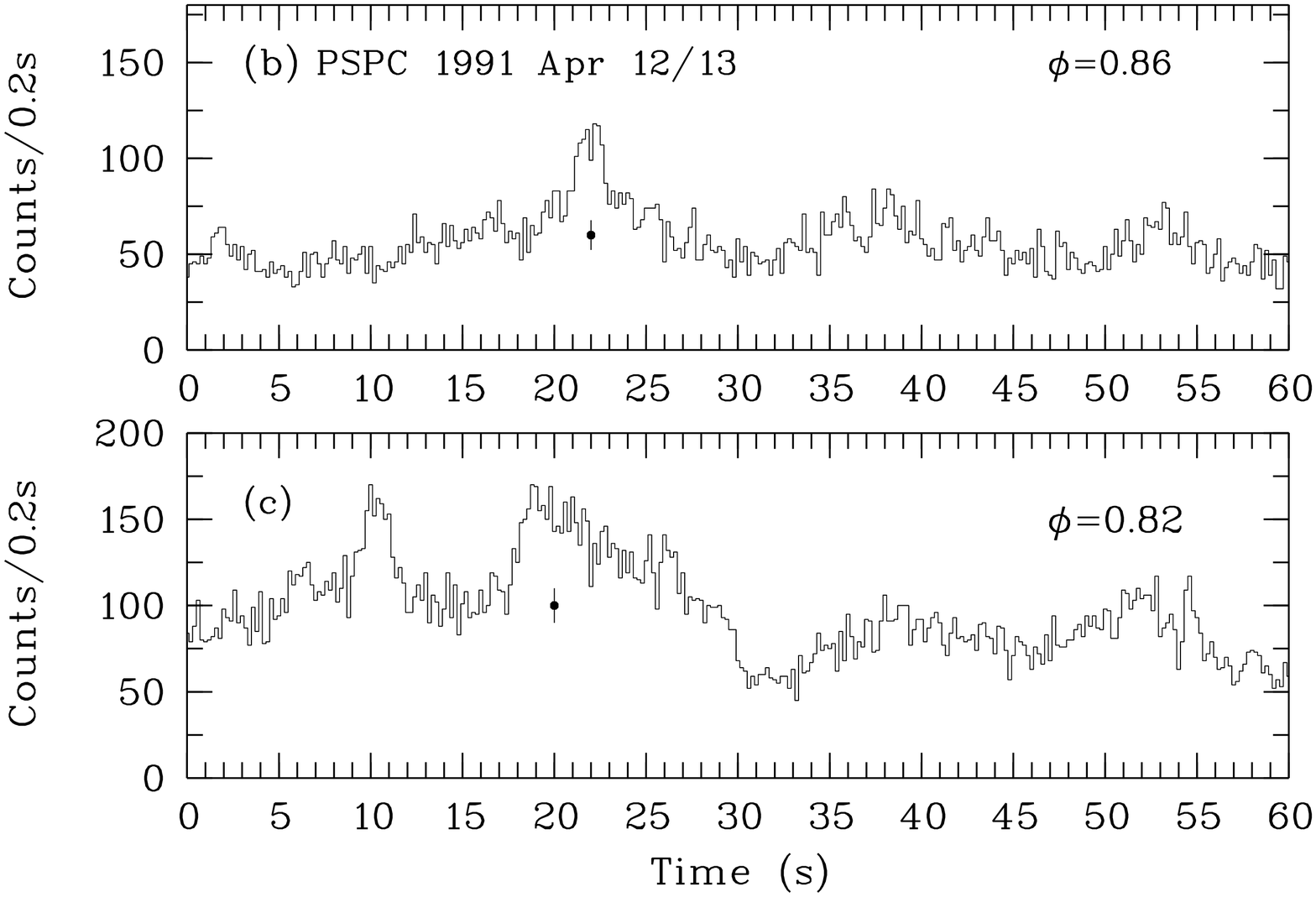}
\includegraphics[width=87mm,angle=0]{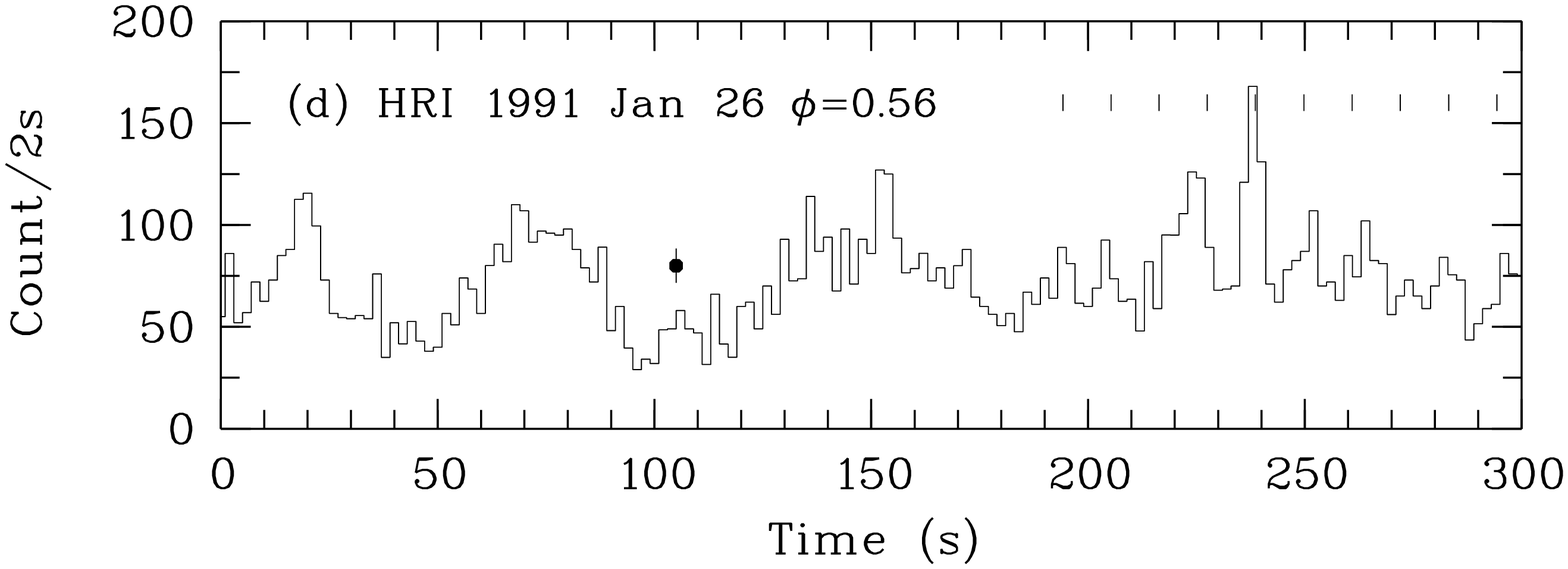}
\caption[chart]{Short-term X-ray variability of AM Her.  \emph{(a)}
PSPC light curve through the orbital dip in the April 1991 high
state. Time bins are 10\,s. For clarity, one observation interval is
shown in green. \emph{(b)} and \emph{(c)} One-minute sections near peak
flux. Time bins are 200\,ms. \emph{(d)} Section of the HRI light
curve in the 1996 high state. Time bins are 2\,s. For all panels, the
typical statistical uncertainty is indicated by the filled circle with
error bars.}
\label{fig:lc}
\end{figure}

The question whether the observed ultraviolet flux represents the
Rayleigh-Jeans tail of the soft X-ray quasi-blackbody emission has
been debated since the discovery of \amher\ as a soft X-ray source
\citep{tuohyetal78,tuohyetal81}. Important steps towards an answer are
the discussion of the energy balance by
\citet{gaensickeetal95,gaensickeetal98} and the detection of rapid
fluctuations in the far-ultraviolet flux by \citet{greeleyetal99},
which lead to the following picture. The white dwarf in \amher\
possesses a large UV-bright spot that covers about 10\% of the white
dwarf surface (full opening angle $70^\circ$, radius
$\ten{5}{8}$\,cm). In the high state, it is heated to a central
temperature of about 50000\,K by bremsstrahlung and cyclotron emission
from the hot post-shock plasma located some $10^8$\,cm above the
surface \citep{gaensickeetal95}. The rise from 50000\,K to the
fluctuating \mbox{300000-K} spotlets of the soft X-ray emission occurs
within a much more restricted region and is largely due to irradiation
by soft X-rays. The energy release in soft X-rays exceeds that in
bremsstrahlung by an order of magnitude, but the emission occurs at
low height. The impact of a dense blob of matter compresses the
photospheric plasma in the respective flux tube and radiative heating
elevates the photosphere in the immediate surroundings creating a
mound \citep{litchfieldking90} of up to several $10^6$\,cm
height. Decompression of a previously mass-loaded flux tube produces a
splash that may reach $10^7$\,cm. The hot matter elevated by these
effects may irradiate a spot of about $10^8$\,cm radius. We identify
this emission with the rapidly fluctuating 'flare-minus-nonflare'
far-ultraviolet component of \citet{greeleyetal99}, which has a mean
high-state bright-phase spectral flux at 930\AA\ of
$\ten{1.6}{-13}$\,\ergsa\ \citep[][see his Figs.~7, 9, and
10]{greeleyetal99}. Corrected for extinction, this number rises to
$\ten{1.85}{-13}$\,\ergsa. The 'flare-minus-nonflare' component
extends to beyond 1800\AA\ and contains temperatures as low as about
5\,eV.

We quote two combinations of three blackbodies each that both match
the PSPC spectrum \emph{as well as} the far-ultraviolet spectral slope
and flux at 930\AA\ and embrace the plausible range of parameters: (i)
a combination of 27.2\,eV, 15\,eV and 5\,eV (Fig.~\ref{fig:meanspec},
top panel) has a total blackbody flux of
$F_\mathrm{bb}=\ten{3.6}{-9}$\,\ergs\ and (ii) the less well fitting
combination of 35\,eV, 22\,eV, and 5\,eV has
$F_\mathrm{bb}=\ten{5.5}{-9}$\,\ergs. The dominant contributors to the
PSPC spectrum and to $F_\mathrm{bb}$ are the components with 27\,eV
and 22\,eV, respectively. We conclude that, although there is some
freedom, the total high-state blackbody flux is confined to
$F_\mathrm{bb}=\ten{(4.5\pm1.5)}{-9}$\,\ergs, with an error that
encompasses also the variations betweeen individual high states
\citep{hessmanetal00}.

\begin{figure*}[t]
\includegraphics[width=60mm,angle=0]{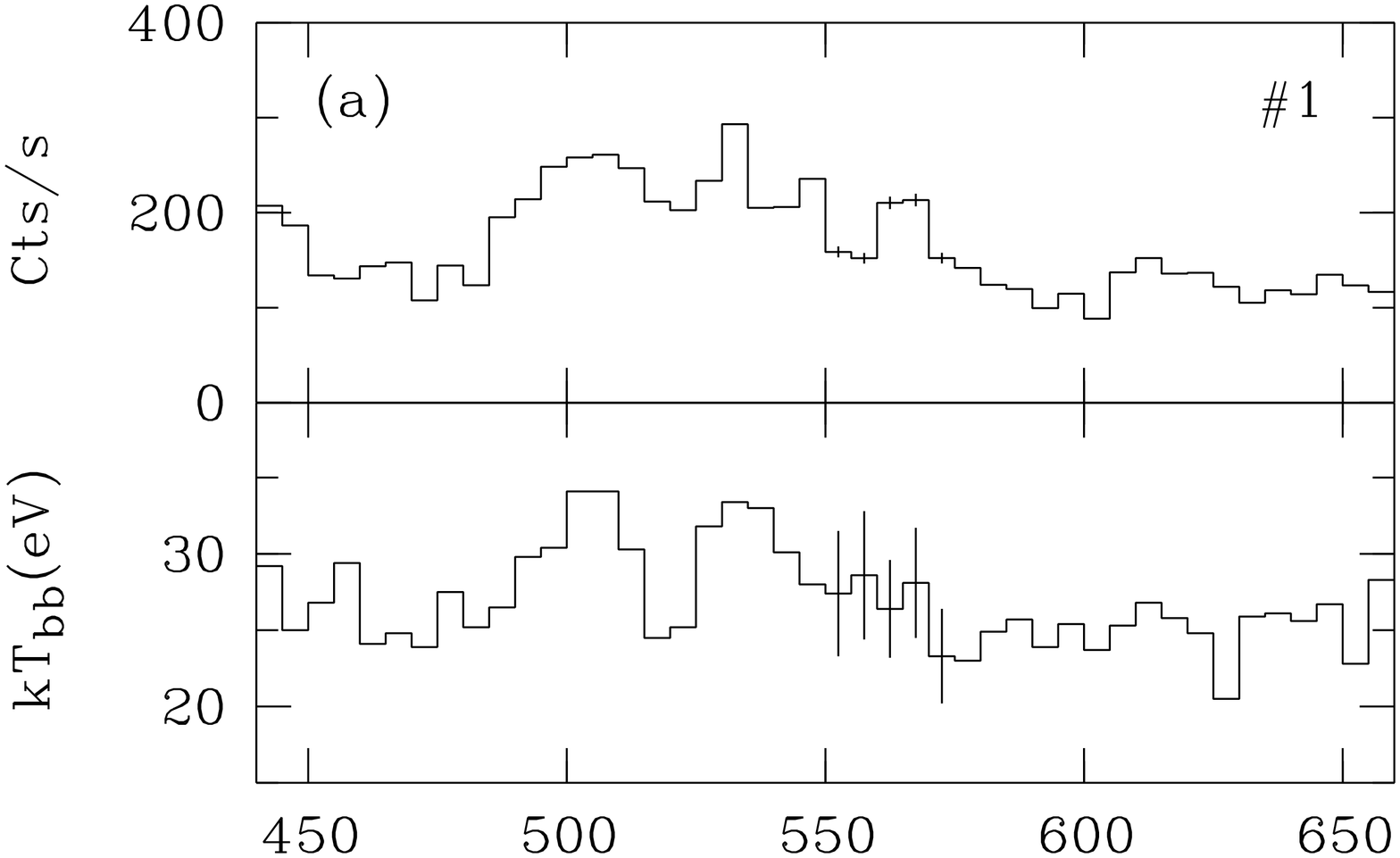}
\hspace*{1.8mm}
\includegraphics[width=57mm,angle=0]{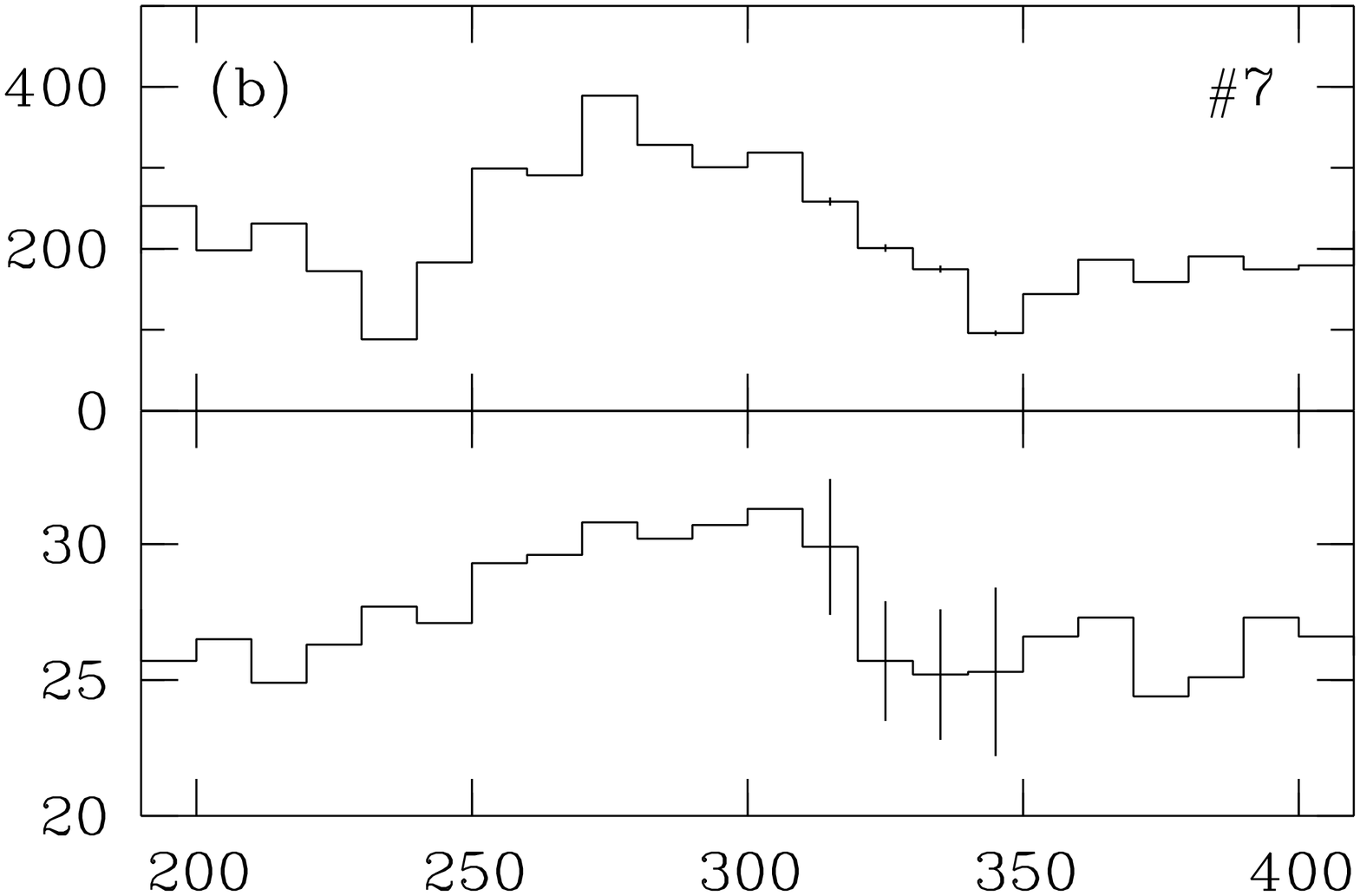}
\hfill
\includegraphics[width=57mm,angle=0]{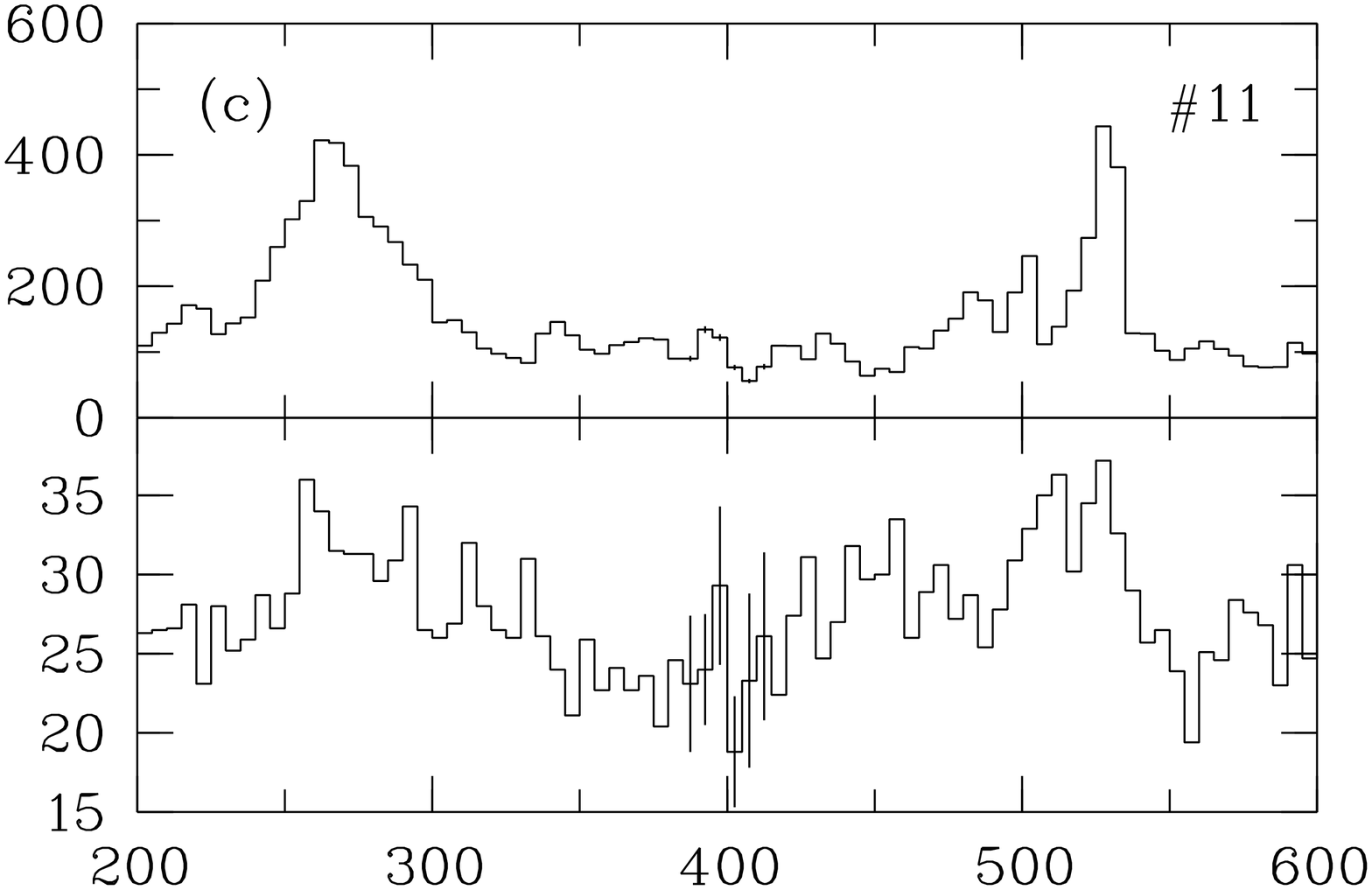}

\vspace*{1mm}
\includegraphics[width=60mm,angle=0]{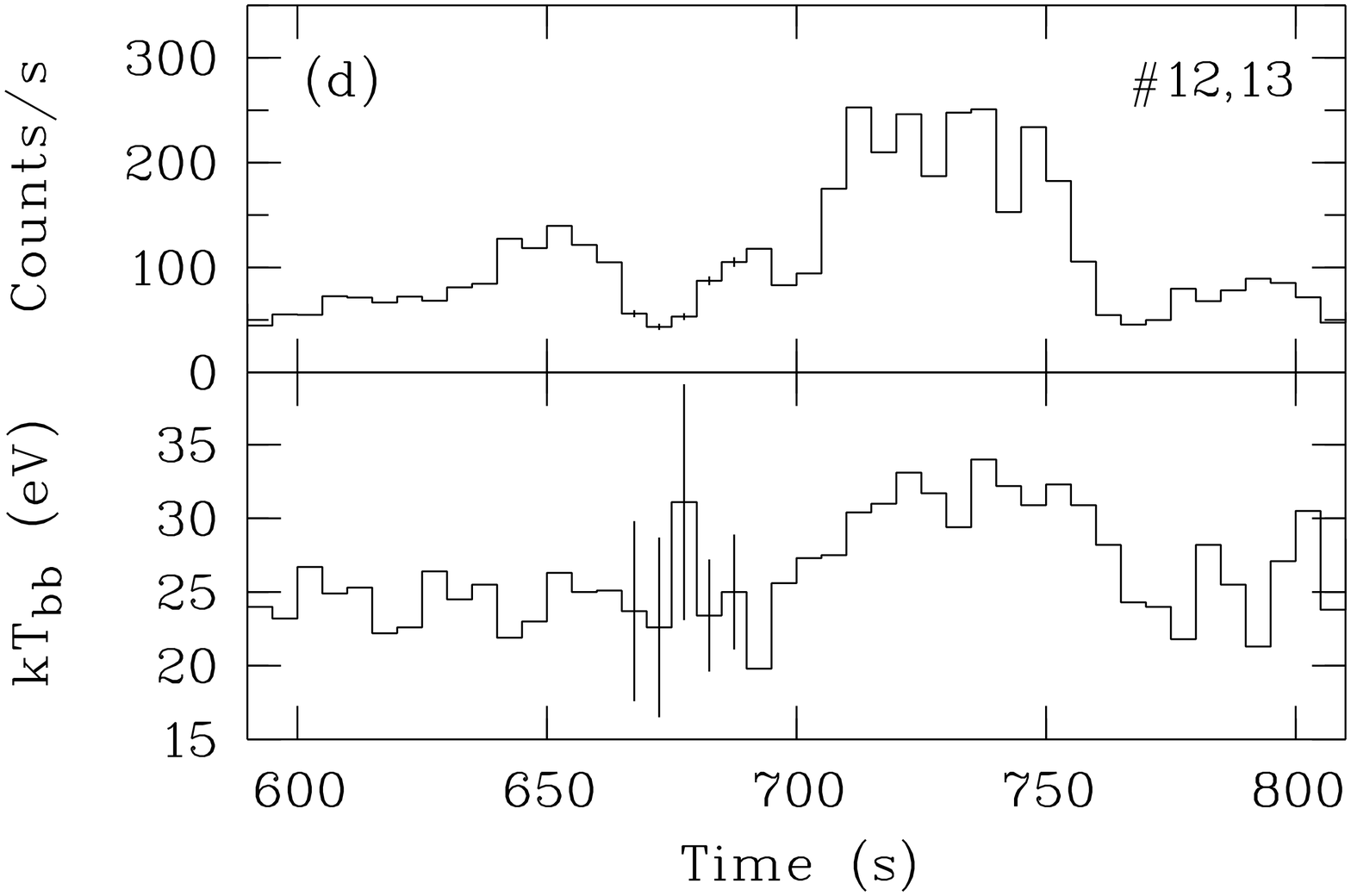}
\hspace*{1.8mm}
\includegraphics[width=57mm,angle=0]{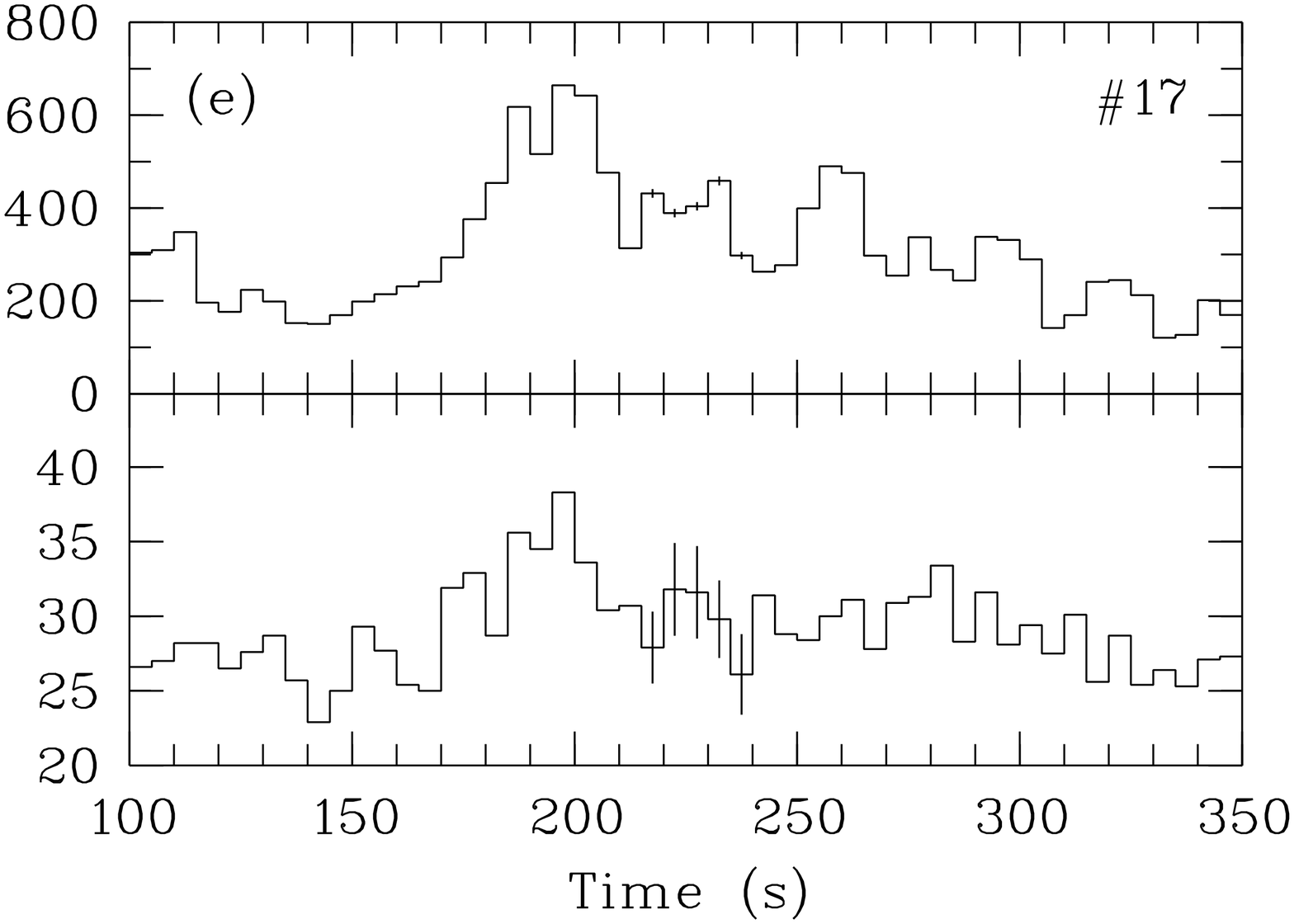}
\hfill
\includegraphics[width=57mm,angle=0]{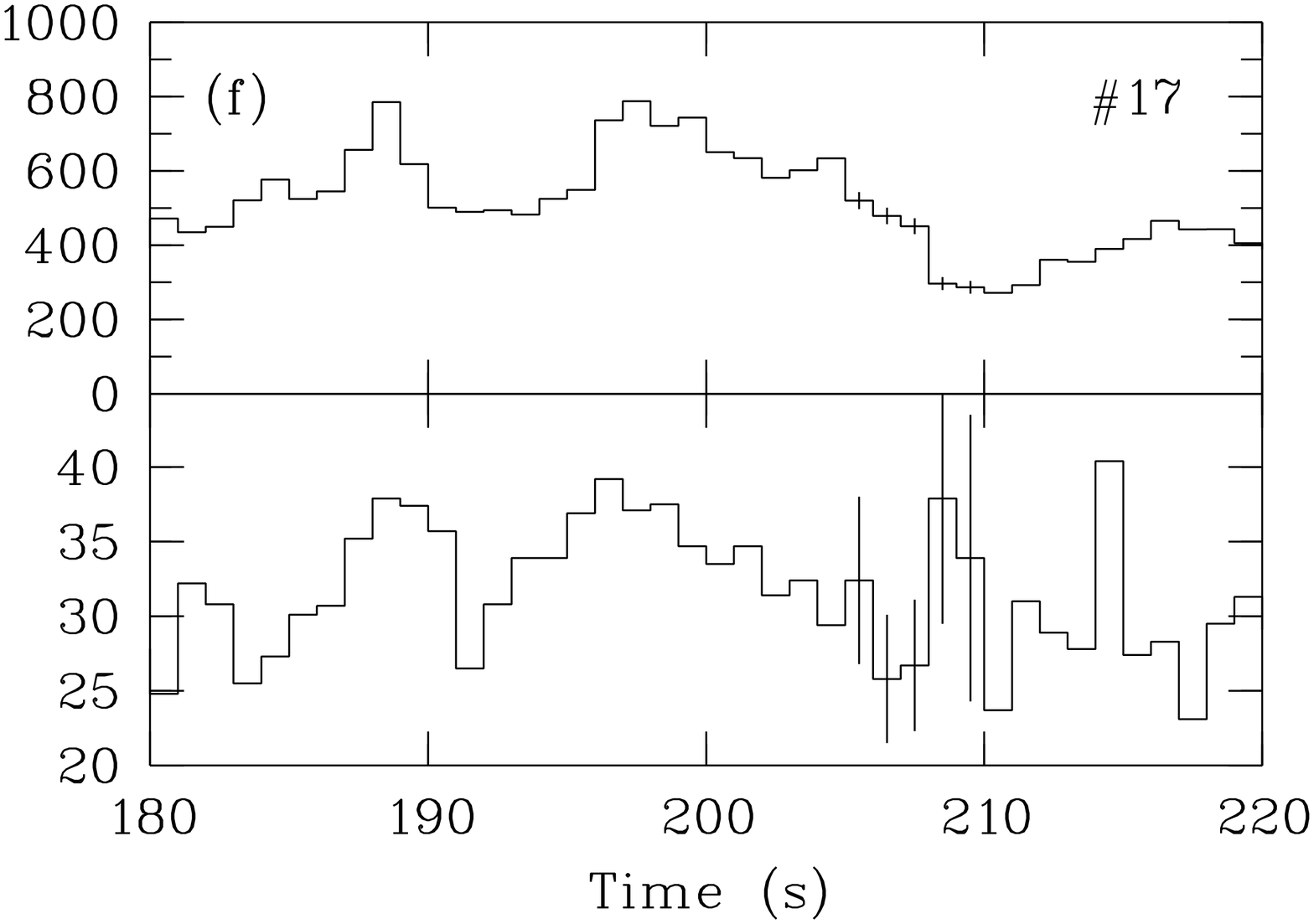}
\caption[chart]{Short-term spectral variability of AM Herculis through
individual flares in the April 1991 high state observation. Plotted is
the PSPC count rate and the fitted blackbody temperature vs. time with
a time resolution of 10\,s in panel \emph{b}, 5\,s in panels \emph{a},
\emph{c}, \emph{d}, and \emph{e}, and 1\,s in panel \emph{f}. Times
start at the beginning of the observation interval (OBI) indicated in
the figures. }
\label{fig:stv}
\end{figure*}

\subsection{The \euve\ and \chandra\ LETG spectra}
\label{sec:chandra}

The mean \euve\ and and the \chandra\ LETG spectra\footnote{Average of
the EUVE nighttime spectra 9309231757N, 9309260306N, and 9503081219N,
and the \chandra\ LETG+HRCS observation 6561 (see
Table~\ref{tab:log}).}  agree within their uncertainties and the PSPC
spectrum, although poorly resolved, is entirely consistent with the
\chandra\ LETG spectrum as demonstrated by \citet{beuermann08}.
Single-blackbody fits to the individual spectra have yielded
k$T_\mathrm{bb}=19.0-22.8$\,eV
for \euve\ \citep{paerelsetal96,christian00} and k$T_\mathrm{bb}=31.7$\,eV 
for \chandra\ \citep{trill06}, with the PSPC temperature falling in
between. The low \euve\ temperature may result from the limited
wavelength range, over which the fit was performed, and to a flux
deficit relative to the LETG spectrum between 70 and 80\AA, where the
background is high. The single-blackbody fit to the \chandra\ LETG
spectrum is poor and a second component with a lower temperature is
required. Surprisingly, the simple model of two blackbodies plus a
high-energy thermal component suffices to obtain an adequate fit over
the entire wavelength range from $\sim20$\AA\ to the interstellar
cutoff at $\sim125$\AA\ if one averages over the emission and
absorption features \citep{trill06}. Trill's best-fit parameters are
$N_\mathrm{H}=\ten{8.8}{19}$\,\atoms, k$T_1=35.9$\,eV, and
k$T_2=15.3$\,eV, of which the latter is uncertain being kept from
running to a lower value by an ultraviolet constraint that is less
restrictive than ours.  If we use our constraint, the total energy
flux stays within the range quoted above. The detailed spectral
structure of the \chandra\ LETG spectrum is difficult to
interprete. Both, a well exposed spectrum and improved model
spectra of irradiated atmospheres are needed to proceed.

\subsection{Luminosity and accretion rate}

With some additional assumptions, the soft X-ray flux allows an
estimate of the accretion rate. The concept of soft X-ray emitting
matter above the mean photospheric level suggests that emission occurs
more or less uniformly into half space \citep[see also][]{heiseetal85}
and $L_\mathrm{bb} =\eta\,d^2 F_\mathrm{bb}$ with $\eta\simeq 2\pi$,
$F_\mathrm{bb}=\ten{(4.5\pm1.5)}{-9}$\,ergs, and $d\simeq80$\,pc
\citep{thorstensen03,beuermann06}.  In addition, the accretion-induced
luminosity $L_\mathrm{acc}$ includes the cyclotron emission with
$F_\mathrm{cyc}\simeq\ten{1.1}{-10}$\,\ergs\ and $\eta\simeq 2\pi$,
and the emission from the accretion stream with
$F_\mathrm{str}\simeq\ten{1.7}{-10}$\,\ergs\ and $\eta$ probably also
near $4\pi$. The bolometric hard X-ray flux in the high state of
\amher\ is estimated at $\ten{2.4}{-10}$\,\ergs\ by
\citet{ishidaetal97} from ASCA data, at more than
$\ten{2.3}{-10}$\,\ergs\ by \citet{beardmoreetal95} from the Ginga
spectrum (the number refers to 2--30\,keV in a moderately high state
with $V=13.7$), and at $\ten{6.4}{-10}$\,\ergs\ by \citet{christian00}
using RXTE data extending to 100\,keV. The hard X-ray geometry factor
is $\eta\simeq 3.3\pi$, which accounts for the reflection albedo
\citep{vanteeselingetal96}. The accretion luminosity then is
$L_\mathrm{acc}=\mathrm{G}M\dot M/R\simeq\ten{(2.2\pm0.8)}{33}$\,\erg,
with $M\simeq 0.78$\,\msun\ and \teff=19800\,K \citep{gaensickeetal06}
the mass and effective temperature of the white dwarf, and
$R\simeq\ten{7.5}{8}$\,cm its radius for a \citet{wood95} model with a
thick hydrogen envelope. The implied accretion rate is $\dot M=
\ten{(1.6\pm0.6)}{16}$\,\gs=$\ten{(2.6\pm0.9)}{-10}$\,\msunyr. The
luminosity ratio is
$L_\mathrm{bb}/(L_\mathrm{hx}+L_\mathrm{cyc})\simeq4\pm2$.

\subsection{X-ray light curves and temporal variability}

On short time scales, the bright phase is characterized by enormous
flaring, which \citet{hameuryking88} have described as the superposition
of an average of 15 accretion-induced flares occurring simultaneously
within the accretion spot. The orbital dip between $\phi\simeq 0.0$
and 0.2 occurs when much of the spot is selfeclipsed by the body of
the rotating white dwarf. The folded orbital light curves of hard
X-rays ($E\ga0.5$\,keV) in the high and low states are shown by
\citet[][see their Figs. 4a and 5a]{gaensickeetal95} and the soft X-ray light
curves of the individual observation intervals (OBIs) in the April
1991 high state by \citet{ramsayetal96}.

Figure~\ref{fig:lc} illustrates the large fluctuations that occur in
the soft X-ray flux of the April 1991 high state for apparent photon
energies $E<0.6$\,keV (channels 10--60). Contrary to earlier
observations \citep[e.g.][]{tuohyetal81} and contrary to the present
HRI data, flaring in the April 1991 PSPC observations does not vanish
entirely during the orbital minimum implying that the accretion spot
rotates just barely out of view (Fig.~\ref{fig:lc}, panel
\emph{a}). While it is possible that a fraction of the residual flux
around $\phi=0.1$ is from the second pole
\citep{tuohyetal81,heiseetal85}, there is no evidence for such an
origin. Substantial variability is known to exist on all time scales
down to \mbox{below 1\,s}. Quasi-periodic X-ray pulsations have been
reported previously \citep[e.g.][]{tuohyetal81,stellaetal86} and quite
persistent pulsations have been seen in the optical cyclotron flux
\citep{bonnetbidaudetal91}, but most of what we see in the \rosat\
data is aperiodic and only short data trains are reminiscent of
quasi-periodicity (e.g. panel \emph{d}). The pulsations of about 1 min
length in panel \emph{d} seem to have a substructure that is around
10\,s in the last pulse (tick marks) and the brightest flares in the
April 1991 PSPC observation (panels \emph{b} and \emph{c}) show
significant variability in subsequent 200\,ms bins.  Thus, the
accretion flow contains structure on length scales from
$L>10^{10}$\,cm corresponding to the $>20$-s fluctuations modeled by
\citet{hameuryking88} down to significant density variations over
scales as short as $10^{8}$\,cm (200\,ms).  We have calculated
autocorrelation functions (not shown), which confirm this picture
showing rapid initial decays, longer tails, and, in part,
cosine-exponential shapes with no preferred periodicity.
The hard X-ray flux ($E>0.6$\,keV) in the high state shows much less
variability than the soft X-rays. We searched for correlated
variations, but found no convincing example, confirming the result of
\citet{stellaetal86}.

\begin{figure}[t]
\includegraphics[width=87mm,angle=0]{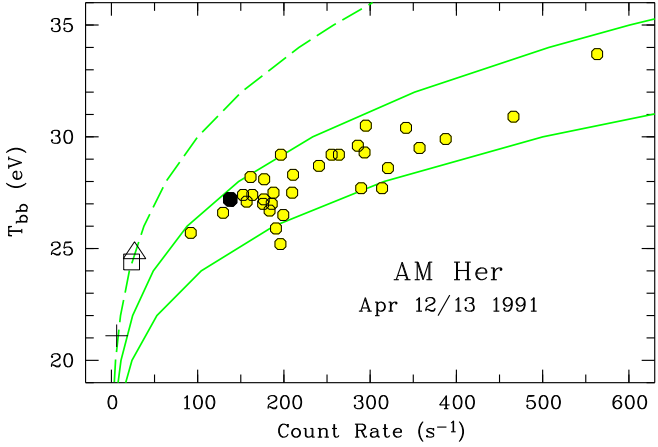}
\caption[chart]{Best-fit blackbody temperature \tbb\ against the PSPC
count rate with time bins of 20\,s for the 13 April 1991 high state
(OBI\,17, open circles). Also shown are the parameters for the mean of
the entire bright phase ({\large $\bullet$}), the flares during the
egress from the dip ($\Box$), the flares in the orbital dip ({\large
$\triangle$}), and the orbital dip without flares ({\large +}). The
curves represent the PSPC count rate vs. \tbb\ dependencies for
different projected blackbody emitting areas on the white dwarf.}
\label{fig:20s}
\end{figure}

\subsection{Short-term spectral variability}

The zoo of temporal structure seen in the X-ray light curves imply a
stream of matter arriving at the white dwarf surface that is highly
structured in space and time. It is generally accepted that the high
density in most individual packets of matter allows them to penetrate
below the photosphere and heat it from below to a temperature that
averages about 25\,eV \citep{kuijperspringle82,franketal88}. To cite a
popular picture, the spot looks like a puddle in a heavy rain. The
stochastic heating and cooling of the individual segments of the spot
has never been observed directly, but the superior statistical
accuracy of the PSPC data combined with its (though moderate) energy
resolution allows the picture to be verified, at least in part.

\begin{figure}[t]
\includegraphics[width=87mm,angle=0]{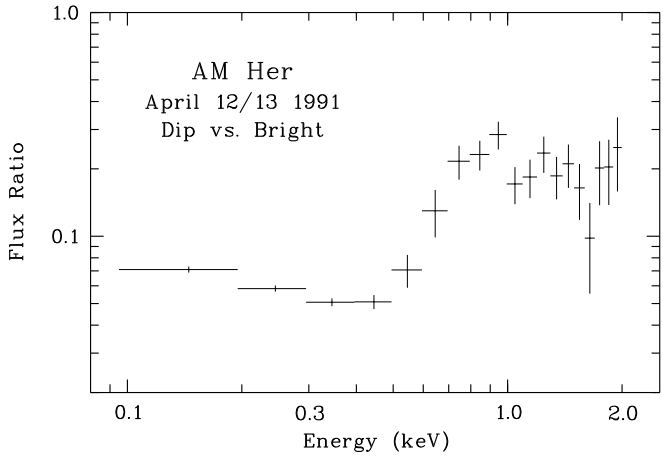}
\caption[chart]{Ratio of the count rate spectra for the bright
phase and the orbital dip of the April 1991 high state
observation.}
\label{fig:dipratio}
\end{figure}

\citet{ramsayetal96} have shown that the blackbody temperature in the
April 1991 PSPC data rises as a function of count rate or that heating
increases with the accretion rate $\dot M$. Ideally, one would like to
search for spectral variability as a function of time with time bins
that resolve the rapid variations of $\dot M$.
With the PSPC, we have established spectral variability in 5\,s bins
and in the largest flare, we can go down to 1\,s. For all
spectral fits discussed in this Section, we have adopted a single
blackbody plus the hard bremsstrahlung component, both absorbed by a
column density fixed at the best-fit value for the mean bright-phase
spectrum. This reduction in freedom is necessary,
because of the reduced statistics of the short exposures and the
strong correlation between the fit parameters \ktbb\ and \nh. In
Fig.~\ref{fig:stv}, we show the fitted blackbody temperature and the
PSPC count rate through several flares. Panel \emph{b} has 10\,s bins,
panels \emph{a}, \emph{c}, \emph {d}, and \emph{e} have 5\,s bins, and
panel \emph{f} shows the brightest section of the flare in panel
\emph{e} at a resolution of 1\,s. The correlated increases of
temperature and count rate are direct proof that localized sections of
the accretion spot are heated by the infall of packets of matter. A
one-to-one correspondence is neither observed nor is it expected,
because the observed temperature averages over surface elements that
heat up and cool independently, and the temperature increments differ
if a given mass flux is concentrated in a single surface element or
distributed over the spot. Fig.~\ref{fig:20s} depicts the results on
the correlation of blackbody temperature vs. count rate for time
intervals of 20\,s near peak count rate (OBI\,17, open circles). Also
included are the parameters of the mean bright-phase spectrum (solid
circle), the rise from the orbital dip (open square), the flares in
the dip (open triangle), and the dip spectrum outside flares ({\large
$+$}). The green curves denote the temperature vs. count rate relation
for blackbody emission from a projected area $A$ on the surface of the
white dwarf absorbed by a column density
$N_\mathrm{H}=\ten{6.3}{19}$\,\atoms. They refer to a distance of
80\,pc and emitting areas $A=\ten{3.4}{14}$\,cm$^2$ (dashed curve),
$\ten{7.9}{14}$\,cm$^2$ (upper solid curve), and
$\ten{1.69}{15}$\,cm$^2$ (lower solid curve). The mean spectrum (solid
circle) corresponds to an emitting area of $\ten{8.7}{14}$\,cm$^2$,
which expands in the flares of OBI\,17 up to
$\ten{2}{15}$\,cm$^2$. During the orbital dip, about 85\% of the soft
X-ray emitting area disappear behind the limb of the white dwarf.

\subsection{Nature of the orbital dip}

The fact that flaring ceases more or less completely in the dip
suggests that we see some kind of sheet lightning caused by hidden
sources behind the limb. This picture is confirmed by the
wavelength-dependent ratio of the PSPC spectra for the dip and the
bright phase (Fig.~\ref{fig:dipratio}).  The bremsstrahlung flux that
dominates the PSPC spectrum at $E>0.6$\,keV is reduced by a factor of
only five, whereas the blackbody flux drops by as much as a factor of
20. This spectral behavior follows naturally from the
three-dimensional geometry of the emitting region with the heated
photosphere disappearing more completely behind the limb than the
vertically extended post-shock region. The dip spectrum at
$E<0.5$\,keV is decidedly softer than that of the bright phase
(Fig.~\ref{fig:dipratio}, \Large{$+$}\normalsize\ and
\Large{$\bullet$}\normalsize\ in Fig.~\ref{fig:20s}), a result that
confirms earlier reports by \citet{paerelsetal96} and
\citet{ramsayetal96}. The softer low-energy spectrum in the dip
reflects the lower temperature in the outer parts of the spot.

\begin{figure}[t]
\includegraphics[width=87mm,angle=0]{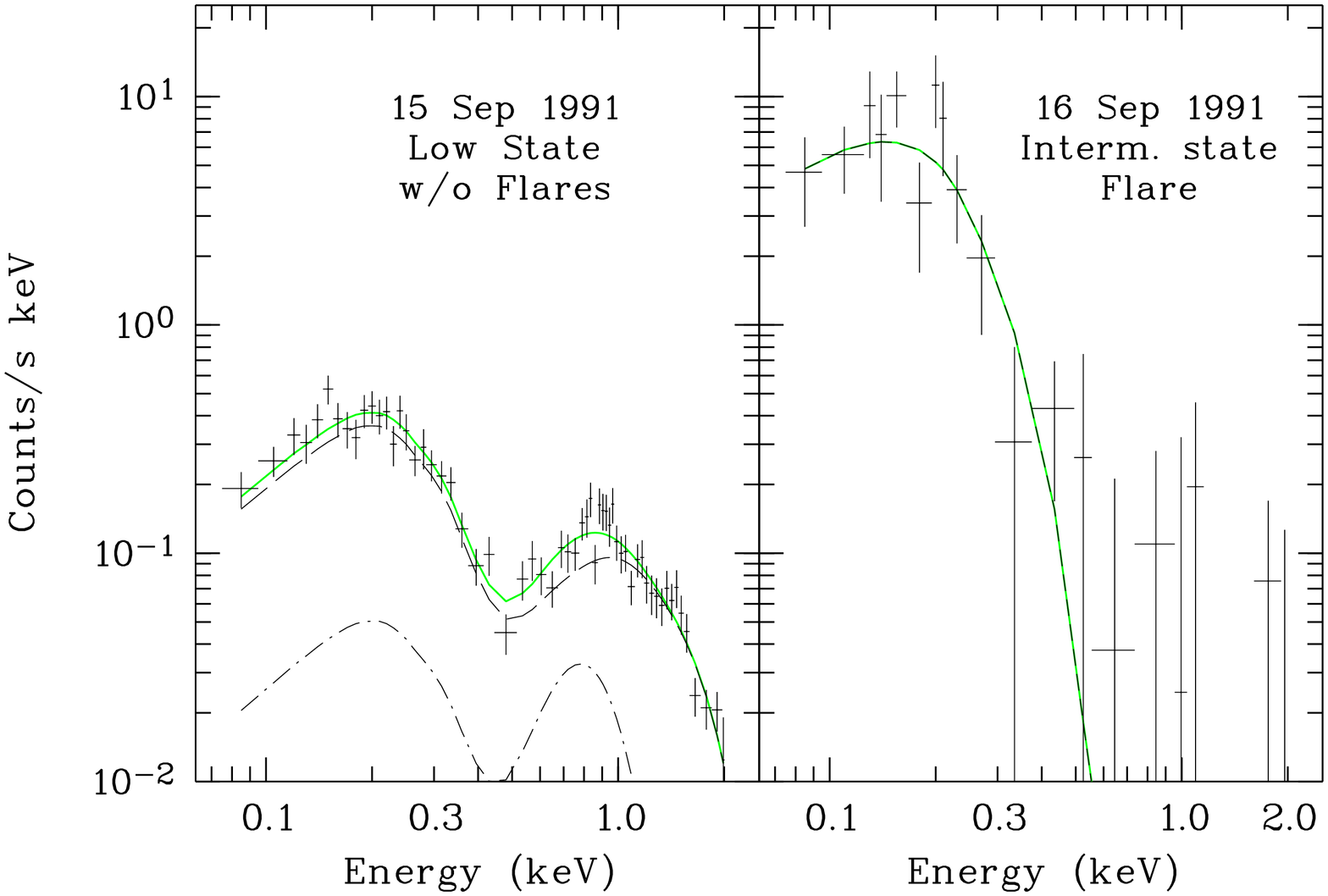}
\caption[chart]{PSPC spectra of \amher\ in the September 1991
low/intermediate state. {\it Left: } 15 September 1991, low state,
bright phase outside flare.  {\it Right: } 16 September 1991, flare
minus non-flare count rate spectrum for single soft X-ray dominated
event. The signature of the curves is as in Fig.~1.}
\label{fig:low}
\end{figure}

\subsection{The September 1991 intermediate and low state}

Following the mid-April 1991 flux maximum, \amher\ declined into a low
state. It was observed with \rosat\ on 15/16 September 1991
(JD\,2448515/6) before it reached minimum visual flux (see Fig.~1 of
Hessman et al. 2000 for the AAVSO light curve). On 15 September the
X-ray flux was very low and only one substantial flare was
observed. The bright-phase spectrum with the flare omitted had a count
rate of only 0.18 cts/s. It can be fitted with the \lya-derived column
density of $\ten{3}{19}$\,\atoms\ and a hard thermal spectrum plus an
0.4\,keV thermal component that accounts for the line emission below
1\,keV (Fig.~\ref{fig:low}, left panel). There is no trace of a
blackbody component, which implies a count rate below
0.05\,cts\,s$^{-1}$, more than a factor of 2000 lower than in the high
state. If the emitting area is the same as in the high state, this
factor limits the low-state blackbody temperature to
\mbox{k$T_\mathrm{bb,low}\la 12$\,eV} (based on the upper solid green
curve in Fig.~\ref{fig:20s}). The flare on the same day and the
generally enhanced flux level on 16 September were also characterized
by a hard spectrum, but on 16 September (HJD2448515.6197) a single
large soft X-ray dominated event with an excess count rate of
1.8\,cts\,s$^{-1}$ occurred. Fig.~\ref{fig:low} (right panel) shows
the spectrum of 88\,s of flare exposure minus the enhanced non-flare
emission underlying and preceding the flare. This difference spectrum
can be fitted by a pure blackbody with a temperature of 21\,eV
(18\,eV) for a cold absorber with
$N_\mathrm{H}=\ten{3}{19}(\ten{6}{19})$\,\atoms. The energy released
in the flare is $\Delta E\simeq \ten{1.1}{33}$\,erg, created by the
infall of about $\Delta M\simeq 10^{16}$\,g. The energy flux in the
high state is larger by a factor of 75 and can be provided by, e.g.,
15 flares occurring simultaneously \citep{hameuryking88} that are each
larger by a factor of five. Our observations suggest that the lack of
soft X-ray emission in the low/intermediate state is due to the rarity
of such flares and not to continued flaring at a temperature too low
for detection with the PSPC. In summary, the low state is
characterized by a tenuous (and occasionally subsiding) flow of
accreting matter that goes through a strong shock above the white
dwarf surface and cools by optically thin thermal X-ray emission and
cyclotron radiation. The intermediate state may be defined by the
occurrence of the first dense packets of matter that penetrate the
photosphere and the high state by the dominance of blobby
accretion. Whether the dense blobs form in the magnetosphere or
already in the L$_1$ nozzle is still an unsolved question.

\section{Summary}

We have analyzed the \rosat\ PSPC bright-phase spectrum in the high
state of \amher\ using an improved detector response matrix. The soft
part of the PSPC spectrum is consistent with a single blackbody of
k$T_\mathrm{bb}=27.4$\,eV and a total blackbody flux of
$F_\mathrm{bb}=\ten{2.6}{-9}$\,\ergs. Assuming a range of temperatures
and considering the limit set by the variable component of the
far-ultraviolet flux \citep{greeleyetal99}, we derive the flux of a
combination of blackbodies that account for the soft X-ray \emph{and}
the far-ultraviolet flux and includes any blackbody that may hide in
the unobservable part of the Lyman continuum:
$F_\mathrm{bb}=\ten{(4.5\pm1.5)}{-9}$\,\ergs. For a distance of
80\,pc, the implied accretion luminosity of \amher\ is
$L_\mathrm{acc}=\ten{(2.1\pm0.7)}{33}$\,\erg\ and the best estimate of
the accretion rate is $\dot M=\ten{(2.4\pm0.8)}{-10}$\,\msunyr. The
spectral parameters reported here are consistent also with the
bright-phase high-state spectra of \amher\ observed with \euve\ and
the \chandra\ LETG+HRC spectrometers.

We have detected rapid soft X-ray spectral variability through the intense
flares in the April 1991 high state of \amher. In many flares,
correlated variations of the blackbody temperature and the PSPC count
rate provide proof of the heating and cooling of sections of the
accretion spot in response to a locally enhanced accretion rate. The
time scale of these variations can be as short as 200\,ms, but even at
peak count rate we can detect spectral variability only for $\Delta
t\ge1$\,s. The Kelvin-Helmholtz time scale of the immediate vicinity
of the impact spot is quite short and the time scale of the observed
spectral variations is essentially that of the $\dot M$
variations. The soft X-ray emitting area of about $10^{15}$\,cm$^2$ is
much smaller than the $\ga10^{17}$\,cm$^2$ of the ultraviolet spot
\citep[e.g.][]{gaensickeetal95,gaensickeetal98,gaensickeetal06}.
In its low state, \amher, emits a hard thermal X-ray spectrum
supported by a flow of tenuous matter. Occasional infall of denser
packets of matter mark the onset of an intermediate state and the
dominance of blobby accretion the high state.

\begin{acknowledgements} The anonymous referee has provided 
constructive comments that improved the presentation of the paper.  We
thank Vadim Burwitz for giving us the mean \chandra\ LETG spectrum of
\amher\ and acknowledge with thanks the observations of
\amher\ from the AAVSO International Database contributed by observers
worldwide. EE and KB thank the National Research Institute (NRIAG),
Helwan, Egypt, for financial support. EE thanks the Deutscher
Akademischer Auslandsdienst (DAAD) for a grant to perform his PhD work
within the Channel-Program at the University of G\"ottingen/Germany.  This
research was in part funded by the DLR under project number
50\,OR\,0501.
\end{acknowledgements}

\bibliographystyle{aa}

\end{document}